\documentclass[%
 aip,
 amsmath,amssymb,
 reprint,%
]{revtex4-1}

\usepackage{graphicx}
\usepackage{caption}
\usepackage{subcaption}
\usepackage{dcolumn}
\usepackage{bm}

\usepackage{listings}
\usepackage[utf8]{inputenc}
\usepackage[T1]{fontenc}
\usepackage{mathptmx}
\usepackage{etoolbox}
\usepackage[most]{tcolorbox}

\usepackage{hyperref}
\hypersetup{
    colorlinks=true,
    linkcolor=blue,
    filecolor=blue,      
    urlcolor=blue,
    citecolor=blue,
    pdftitle={Overleaf Example},
    pdfpagemode=FullScreen,
    }
\usepackage{listings}
\usepackage{color} 
\definecolor{mygreen}{RGB}{28,172,0} 
\definecolor{mylilas}{RGB}{170,55,241}
\usepackage{matlab-prettifier}

\makeatletter
\def\@email#1#2{%
 \endgroup
 \patchcmd{\titleblock@produce}
  {\frontmatter@RRAPformat}
  {\frontmatter@RRAPformat{\produce@RRAP{*#1\href{mailto:#2}{#2}}}\frontmatter@RRAPformat}
  {}{}
}%
\makeatother
\begin{document}

\preprint{AIP/123-QED}

\title{$\mu$TRec: A Muon Trajectory Reconstruction Algorithm for Enhanced Scattering Tomography}
\author{Reshma Ughade}%
\altaffiliation[ ]{Email: rughade@purdue.edu. }
\affiliation{ 
School of Nuclear Engineering, Purdue University, West Lafayette, IN  47907, USA
}%

\author{Stylianos Chatzidakis}%
\affiliation{ 
School of Nuclear Engineering, Purdue University, West Lafayette, IN  47907, USA
}%

\date{\today}

\begin{abstract}

Cosmic ray muons are highly penetrating subatomic particles, making muon tomography a powerful non-invasive technique for imaging dense and large-scale structures. As muons traverse matter, they undergo multiple Coulomb scattering (MCS), deviating from their original trajectory. The degree of scattering is dependent on the material's atomic number (Z), enabling the identification and differentiation of materials. Muon tomography has found applications in diverse fields such as geology, archaeology, and nuclear security. Traditionally, reconstruction algorithms such as the Straight-Line Path (SLP) and Point of Closest Approach (PoCA) have been employed for muon imaging. However, these methods assume single scattering and fail to accurately represent the cumulative nature of MCS. This paper introduces a novel reconstruction method, the Muon Trajectory Reconstruction ($\mu$TRec) algorithm, which provides a more accurate approximation of the muon path by incorporating a statistical framework. Based on a Bayesian framework with Gaussian approximations, the proposed $\mu$TRec algorithm reconstructs the curved trajectories of muons as they traverse matter, incorporating both multiple Coulomb scattering and linear energy loss to accurately map scattering angles along the path. The proposed algorithm is applied to simulate imaging of dry casks used for spent nuclear fuel storage considering the horizontal orientation. Four cask loading conditions are evaluated: fully loaded, one row missing, one assembly missing, and half assembly missing. Simulations are performed using $10^5$ and $10^6$ muons to assess imaging performance. The results demonstrate improved accuracy and resolution in identifying missing assemblies compared to conventional methods. It is observed that the $\mu$TRec algorithm exhibits markedly superior performance over the classical PoCA method achieving respective improvements of \textbf{122\%} in signal-to-noise ratio (SNR), \textbf{35\%} in contrast-to-noise ratio (CNR), and \textbf{201\%} in detection power (DP) for the case of one missing fuel assembly with a muon flux of $10^6$ and a voxel size of 5~cm. Furthermore, $\mu$TRec supports high-resolution reconstruction with voxel sizes as small as 1~cm which enables the accurate localization of structural components such as the 2.5~cm thick steel canister. Notably, $\mu$TRec is also capable of reliably detecting a single missing fuel assembly at a muon flux as low as $10^5$, a task that remains infeasible using PoCA under the same conditions.

\end{abstract}

\maketitle

\section{\label{sec:level1} Introduction}

Conventional imaging methods primarily utilize X-rays, neutrons, and more recently protons. However, X-rays exhibit limited penetration through dense or heavily shielded structures~\cite{liu_simulation_2016}. Additionally, neutron and proton imaging typically require costly accelerator facilities resulting in increased complexity and operational expenses. In contrast, cosmic ray muons, charged subatomic particles approximately 200 times heavier than electrons provide substantial advantages due to their inherent penetrating capabilities and natural abundance at Earth's surface with an average flux of approximately $10^4$ muons~m$^{-2}$min$^{-1}$ at sea level \cite{bae_gamma-ray_2024}. Muon tomography thus eliminates the need for accelerator-based sources making it cost-effective and practical for imaging thick, dense, and highly shielded structures. Since the pioneering work by George in 1955~\cite{cosmic_rays_1955}, muon tomography has increasingly gained attention, particularly in the past two decades. Today, it is widely applied to image complex and dense objects such as volcanoes~\cite{nagamine_method_1995}, cargo containers~\cite{baesso_toward_2014}, civil infrastructure~\cite{odonnell_muographic_2025}, and nuclear waste repositories~\cite{jenneson_large_2004, thomay_passive_2016}. In nuclear engineering, muon imaging is particularly advantageous for detecting and localizing nuclear fuel debris in damaged reactors~\cite{borozdin_cosmic_2012, miyadera_imaging_2013}. A significant application of muon tomography is the inspection and monitoring of dry storage casks (DSC) that house spent nuclear fuel from reactors. Monitoring DSCs to detect any changes in internal configurations after transportation is essential for effective nuclear waste management. DSCs are engineered with dense shielding materials, typically comprising concrete walls to absorb neutron radiation and steel canisters to attenuate gamma radiation ~\cite{gao_dry_2020, el-samrah_radiation_2022}. Consequently, imaging the interior of these heavily shielded and structurally dense casks presents a considerable challenge, motivating recent research efforts to accurately detect anomalies such as missing fuel assemblies~\cite{poulson_cosmic_2017, bae_momentum_2024, li_muon_2019, chatzidakis_classification_2017, chatzidakis_investigation_2016, park_design_2022, jonkmans_nuclear_2013, chatzidakis_interaction_2016, durham_verification_2018, chatzidakis_analysis_2016}.

Recent developments in muon tomography have been constrained by several factors, including the long data collection times necessary to acquire sufficient muon events, limitations inherent to current reconstruction algorithms in modeling MCS, and the complexities involved in image processing. Muon detection systems provide measurements of muon position and momentum before and after passing through the target object. However, the actual trajectories of muons through dense structures cannot be directly measured and must be estimated through reconstruction algorithms. This estimation is complicated by the intrinsic scattering and energy loss experienced by muons within the material. Present reconstruction techniques typically employ simplistic geometrical approximations, leading to reduced spatial resolution and image blurring. One common approach is the SLP algorithm~\cite{poulson_cosmic_2017}, which neglects scattering and assumes a linear trajectory between the entry and exit points of the muon. Maximum Likelihood Expectation Maximization (MLEM) algorithm assumes a straight line path between entry and exit points and distributes the likelihood of scattering across the voxels intersected by this path. Iteratively, the algorithm refines voxel-wise scattering densities to maximize the likelihood of observing the measured angular deviations~\cite{schultz_statistical_2007, ruta_iterative_2023, vanini_muography_2018, wang_muon_2018, yang_novel_2018, wang_statistical_2009, chaiwongkhot_3d_2022}. While effective in capturing statistical trends, MLEM is computationally intensive and typically demands a high muon flux to achieve accurate reconstruction. Another widely used method, the PoCA algorithm, assumes a single scattering event by calculating the midpoint of the shortest distance between the extrapolated incoming and outgoing muon paths~\cite{schultz_image_2004}. However, this assumption may exclude muon events whose paths do not intersect or midpoints aren't within the reconstruction volume, thereby reducing the number of usable muon trajectories and necessitating longer measurement durations to achieve acceptable image quality. Consequently, there is a critical need for improved reconstruction algorithms capable of accurately modeling muon trajectories, incorporating realistic MCS and energy loss effects. 

In the present study, we address these challenges by proposing a statistical approach to more accurately approximate muon trajectories. We introduce a novel reconstruction algorithm named the  Muon Trajectory Reconstruction ($\mu$TRec), which leverages a Bayesian statistical framework combined with Gaussian approximations of multiple Coulomb scattering~\cite{schultz_statistical_2007, ughade_performance_2023, chatzidakis_generalized_2018, ughade_--fly_2023, ughade_physics-based_2023, 38c69f84edc6412aba4773e561e20650, ughade2023muon}. Unlike conventional methods, $\mu$TRec yields a curved muon trajectory that asymptotically aligns with the measured incoming and outgoing paths, thereby providing a more precise representation of the muon's actual path through the target material. Enhanced trajectory accuracy is expected to significantly improve the resulting image resolution and overall imaging quality. To evaluate the efficacy of the $\mu$TRec algorithm relative to the widely used PoCA method, we conducted simulations for the horizontal orientation of DSC. Four specific scenarios were investigated to assess the sensitivity to internal structural changes within the DSCs: a fully loaded configuration, one row of fuel assemblies missing, a single missing fuel assembly, and a half missing fuel assembly. Initial results obtained using a mono-energetic and unidirectional muon source indicate that the $\mu$TRec method can successfully identify even subtle anomalies such as a half missing fuel assembly with as few as $10^5$ muon events. Additionally, all scenarios were revisited and simulated with a more realistic poly-energetic and multi-directional muon flux generated using a comprehensive muon source model ``Muon Generator''~\cite{chatzidakis_geant4-matlab_2015}. This revision aims to validate and enhance the practical applicability and robustness of the $\mu$TRec algorithm under realistic operational conditions.

\section{Geant4 Modelling}

\subsection{Cosmic ray muon simulations}

\begin{figure}[htbp]
\centerline{\includegraphics[scale = 0.5]{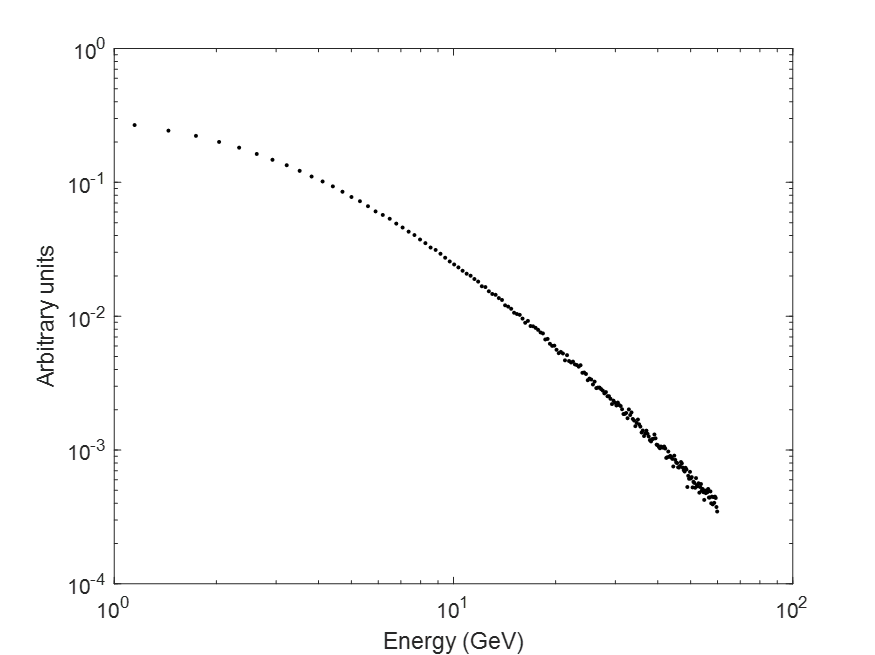}}
\caption{Energy distribution of simulated muon particles in Geant4}
\label{energy}
\end{figure}

\begin{figure}[htbp]
\centerline{\includegraphics[scale = 0.5]{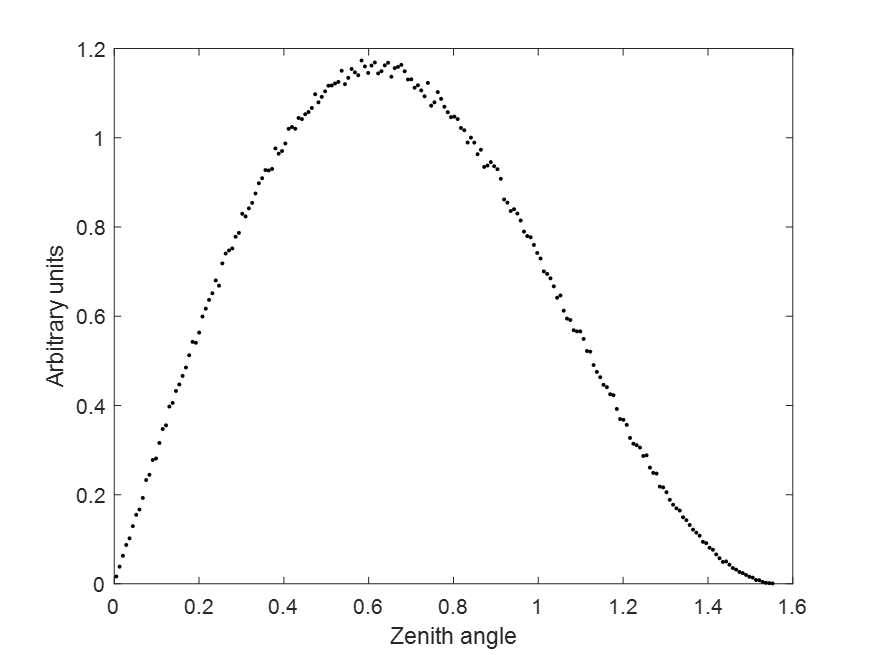}}
\caption{Angular distribution of simulated muon particles in Geant4}
\label{zenith angle}
\end{figure}

In this study, a realistic cosmic-ray muon flux was simulated using a dedicated muon generator~\cite{chatzidakis_geant4-matlab_2015}. The generated muon energy spectrum ranged from 1-60~GeV, while the zenith angles varied between 0$^\circ$ and 90$^\circ$, as illustrated in Figures~\ref{energy} and~\ref{zenith angle}, respectively. This broad energy and angular distribution closely resembles the natural characteristics of cosmic-ray muons observed at Earth's surface, thereby enhancing the reliability and practical relevance of the simulations.

\subsection{Spent nuclear fuel dry cask storage}

\begin{figure}[htbp]
    \centering
    \includegraphics[width=0.9\linewidth]{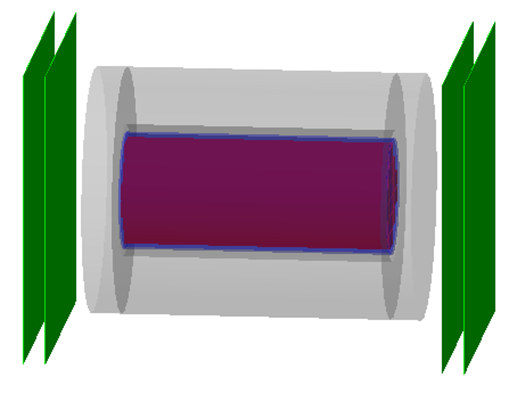}
    \caption{Horizontal orientation of the dry fuel cask with four muon detectors (green) placed at the entrance and exit along the muon trajectory.}
    \label{horizontal}
\end{figure}

\begin{figure}[htbp]
    \centering
    \includegraphics[width=0.7\linewidth]{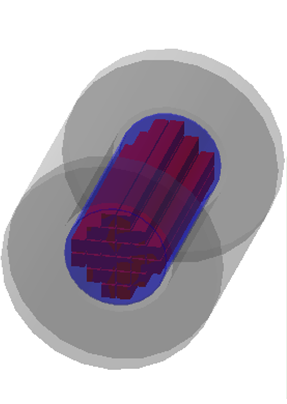}
    \caption{VSC-24 dry fuel cask model showing 24 fuel assemblies (red), enclosed within a steel canister (blue), and surrounded by a cylindrical concrete shielding structure (grey).}
    \label{cask}
\end{figure}

In this study, the VSC-24 dry fuel cask is selected to evaluate the performance of the proposed $\mu$TRec algorithm \cite{EPRI2010SpentFuel}. The cask stores 24 fuel assemblies, each with dimensions of $21 \times 21 \times 361\,\text{cm}^{3}$. These fuel assemblies are enclosed within an annular steel canister (density $7.93\,\text{g/cm}^{3}$) having an inner radius of $77\,\text{cm}$ and an outer radius of $79.5\,\text{cm}$. The steel canister is further surrounded by concrete shielding (density $2.3\,\text{g/cm}^{3}$), which has an inner radius of $89.5\,\text{cm}$ and an outer radius of $167.5\,\text{cm}$. Finally, the dry fuel cask is sealed by a lid composed of a $2.5\,\text{cm}$ thick steel layer and a $41.5\,\text{cm}$ thick concrete layer. The horizontal configuration of the DSC including two incoming and two outgoing muon detectors is illustrated in Figure \ref{horizontal}. A detailed visualization of the VSC-24 DSC including its internal components is shown in Figure \ref{cask}.

To comprehensively assess the capability of the $\mu$TRec algorithm, four representative scenarios reflecting potential changes in internal fuel configuration are analyzed:
\begin{enumerate}
    \item Fully loaded fuel cask,
    \item One complete column of fuel assemblies missing,
    \item Single fuel assembly missing,
    \item Half of one fuel assembly missing.
\end{enumerate}

\begin{figure*}[htbp]
    \centering
    \subfloat[Fully loaded cask.]{%
        \label{4figs-a}%
        \includegraphics[width=0.23\textwidth]{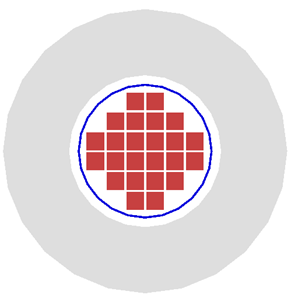}%
    }\hfill
    \subfloat[One column of fuel assemblies missing.]{%
        \label{4figs-b}%
        \includegraphics[width=0.23\textwidth]{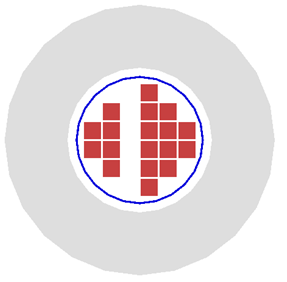}%
    }\hfill
    \subfloat[One fuel assembly missing.]{%
        \label{4figs-c}%
        \includegraphics[width=0.23\textwidth]{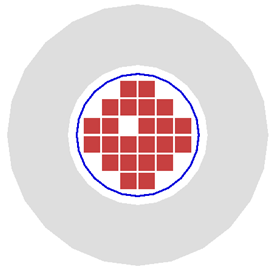}%
    }\hfill
    \subfloat[Half fuel assembly missing.]{%
        \label{4figs-d}%
        \includegraphics[width=0.23\textwidth]{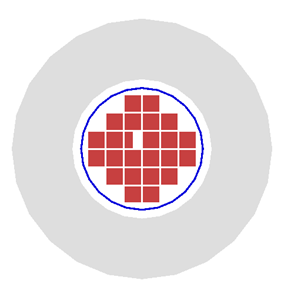}%
    }
    \caption{Illustrations of the four dry fuel cask scenarios for analysis: (a) fully loaded (b) one column of fuel assemblies missing (c) single fuel assembly missing and (d) half fuel assembly missing.}
    \label{4figs}
\end{figure*}

These specific scenarios are visually depicted in Figure~\ref{4figs}.

\subsection{Scintillation detectors} 

Muon position data in this study were obtained using an array of four scintillation detectors. Each detector has dimensions of $400 \times 400 \times 1\,\text{cm}^{3}$. The detectors are arranged horizontally with a total separation of 600\,cm between the first and last detector. Specifically, the first two detectors and the last two detectors each have an inter-detector spacing of $30\,\text{cm}$. For the purposes of this analysis, the detectors are assumed to possess ideal spatial resolution.

\section{Muon Trajectory Estimators}

In this study, the performance of the proposed $\mu$TRec algorithm is evaluated and compared against the widely adopted PoCA algorithm. The methodological frameworks underlying both reconstruction techniques are presented in this section.

\subsection{Point of Closest Approach}

\begin{figure}[h]
    \centering
    \includegraphics[width=220px]{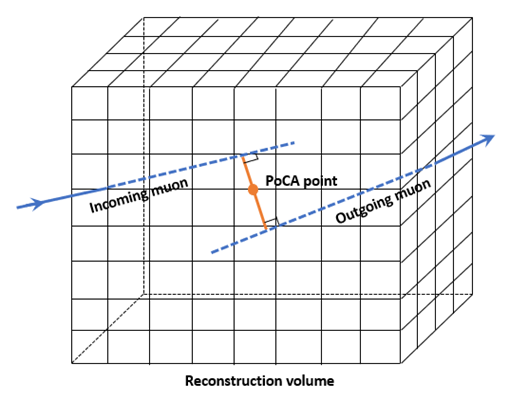}
    \caption{\centering Scattering point for PoCA algorithm.}
    \label{fig:PoCA}
\end{figure}

To reconstruct the scattering locations of muons within an object, the PoCA algorithm adopts a geometrically simplified model. This method operates under the assumption that the total angular deflection a muon experiences while traversing a medium can be attributed to a single discrete scattering event. Unlike more sophisticated models that attempt to capture the cumulative effects of MCS along the muon path, PoCA approximates the entire deviation by assigning a single point to represent the interaction. The implementation begins with the determination of incoming and outgoing muon trajectories based on position measurements collected from two pairs of tracking detectors. It is assumed that the detectors have sufficient resolution to accurately associate each recorded hit with its corresponding muon track. From this data, two vectors are constructed to represent the entry and exit paths. The intersection of these two vectors is considered the scattering point. However, these vectors may not intersect in three-dimensional space. Instead, the PoCA point is computed by identifying the pair of points, one on each trajectory, that are closest to one another in Euclidean distance. This is accomplished using a least-squares minimization of the perpendicular distance between the two lines. The midpoint of this shortest segment is then defined as the scattering point as shown in Figure \ref{fig:PoCA}, and the scattering angle is estimated as the angular difference between the two trajectory vectors.

While PoCA offers computational simplicity and has been widely adopted in muon tomography studies, it inherently neglects the stochastic and distributed nature of MCS. Moreover, its reliance on a single scattering point can lead to inaccuracies, particularly in complex geometries or when the scattering is weak and distributed. Additionally, muons for which the extrapolated entry and exit paths do not converge within the reconstruction volume are often discarded, reducing the effective number of events available for image formation and increasing the data acquisition time required for meaningful results.

\subsection{Generalized Muon Trajectory Estimation}

The trajectory reconstruction algorithm presented in this work builds upon foundational concepts previously developed for estimating the most probable paths of charged particles, such as the method proposed by Schulte et al.~\cite{schulte_maximum_2008}, which focused on proton transport through homogeneous media. While influential, these earlier approaches were limited in scope to lower-energy applications and often required complex evaluations of polynomial ratios to model energy loss. Furthermore, they were not directly applicable to high-energy cosmic ray muons or to scenarios involving heterogeneous, high-$Z$ materials commonly encountered in nuclear engineering and geophysical imaging. To address these challenges, the algorithm developed in this study adopts a generalized statistical framework tailored for cosmic ray muon tomography, following the approach introduced by Chatzidakis et al.~\cite{chatzidakis_generalized_2018, chatzidakis_2016_cosmic}. The method uses a bivariate Gaussian approximation to characterize MCS, incorporating both angular deflections and lateral displacements. Energy loss is treated using the continuous slowing down approximation (CSDA), which is appropriate given the nearly constant energy loss exhibited by minimum ionizing muons over a broad energy range. Unlike conventional techniques that rely on straight-line approximations or assume single-point scattering, this algorithm estimates a continuous, curved trajectory that more accurately reflects the physical interactions experienced by muons as they traverse dense and non-uniform materials. The estimated path asymptotically approaches the measured incoming and outgoing muon trajectories, leading to enhanced spatial resolution and improved image fidelity. 

\subsubsection*{Trajectory Equation}

\begin{figure}[h]
    \centering
    {\includegraphics[width=230px]{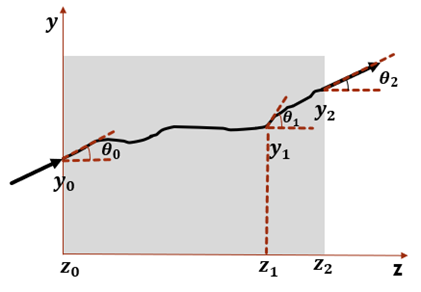}}
    \caption{Illustration of muon scattering occurring within the $y$–$z$ plane.}
    \label{fig:method_graph}
\end{figure}

Given that the lateral (x) and vertical (y) scattering of muons due to multiple Coulomb interactions are statistically independent processes, the trajectory of a muon can be decomposed into two-dimensional projections. This allows the complex three-dimensional trajectory to be reconstructed through separate analyses in the $x$–$z$ and $y$–$z$ planes \cite{ughade_3d_2024}. For clarity let us consider the $y$–$z$ projection as illustrated in Figure~\ref{fig:method_graph}, where the $z$-axis is oriented perpendicular to the detector planes, and the $x$ and $y$ axes lie parallel to the plane of the detectors. In this framework, the position and scattering angle of the muon are known at two detector planes located at depths $z_0$ and $z_2$. These are represented as $(y_0, \theta_0)$ and $(y_2, \theta_2)$, respectively, where $y$ denotes the transverse position and $\theta$ the local scattering angle. The goal is to estimate the position and angle of the muon at an intermediate depth $z_1$, situated between the two detectors. The muon state is represented using a two-dimensional state vector that encapsulates both the transverse displacement and angular deflection at $z$. 

$ \mathbf{Y}  = \begin{pmatrix}
  $y$ \\ \theta
  
\end{pmatrix} 
$

The Gaussian approximation of Y follows a zero-mean bivariate normal distribution.

\begin{equation}
    \mathcal{P}(\mathbf{Y}) \sim \mathcal{N}(\mathbf{0}, \boldsymbol{\Sigma}),
\end{equation}
where $\boldsymbol{\Sigma}$ is the covariance matrix
\begin{equation}
    \boldsymbol{\Sigma} = 
    \begin{bmatrix}
        \sigma_y^2 & \sigma_{y\theta_y} \\
        \sigma_{y\theta_y} & \sigma_{\theta_y}^2
    \end{bmatrix},
\end{equation}

where $\sigma_y^2$ is the scattering displacement variance, $\sigma_\theta^2$ is the scattering angle variance, and $\sigma_{y\theta}^2$ is the covariance between the displacement and the scattering angle. The values of these quantities can be estimated from simulation, theory, or from actual muon measurements. This leads to the general form

\begin{equation}
P(\mathbf{Y}) = \frac{1}{2\pi \sqrt{|\bm{\Sigma}|}} \exp\left(-\frac{1}{2} \mathbf{Y}^T \bm{\Sigma}^{-1} \mathbf{Y} \right),
\end{equation}

Using Bayesian theory, the probability of a muon displaying a specific displacement and angle at $y_1$, given the exit point 2, can be expressed as follows:

\begin{align}\label{eq:3_1}
P(y_1 \vert  y_2) = \frac{P(y_2 \vert  y_1) P(y_1 \vert  y_0)}{P(y_2)} 
\end{align}

Since $P(y_2)$ is a normalization constant, we aim to maximize:

\begin{equation}\label{proportional}
    P(y_1 | y_2) \propto P(y_2 | y_1) P(y_1 | y_0).
\end{equation}

Assuming both terms are Gaussian distributions, we write:

\begin{equation}
    P(y_1 | y_0) = \frac{1}{C_1} \exp\left( -\frac{1}{2} (y_1 - R_0 y_0)^T \Sigma_1^{-1} (y_1 - R_0 y_0) \right),
\end{equation}

\begin{equation}
    P(y_2 | y_1) = \frac{1}{C_2} \exp\left( -\frac{1}{2} (y_2 - R_2 y_1)^T \Sigma_2^{-1} (y_2 - R_2 y_1) \right),
\end{equation}

where $C_1$, $C_2$ are constants and

\begin{equation}\label{Transport_matrices}
    R_0 = \begin{bmatrix} 1 & z_1 - z_0 \\ 0 & 1 \end{bmatrix}, \quad
    R_2 = \begin{bmatrix} 1 & z_2 - z_1 \\ 0 & 1 \end{bmatrix},
\end{equation}

\begin{equation}\label{covariance_matrices}
\Sigma_1 = \begin{pmatrix} \sigma_{y_1}^2 & \sigma_{y_1 \theta_1}^2 \\ \sigma_{y_1 \theta_1}^2 & \sigma_{\theta_1}^2  \end{pmatrix}, \quad
\Sigma_2 = \begin{pmatrix} \sigma_{y_2}^2 & \sigma_{y_2 \theta_2}^2 \\ \sigma_{y_2 \theta_2}^2 & \sigma_{\theta_2}^2  \end{pmatrix}.
\end{equation}

Taking the logarithm of equation \ref{proportional} :

\begin{align}
    \ln P(y_1 | y_2) &= \ln P(y_2 | y_1) + \ln P(y_1 | y_0) + \text{const} \\
    &= -\frac{1}{2} (y_1 - R_0 y_0)^T \Sigma_1^{-1} (y_1 - R_0 y_0) \notag \\
    &\quad -\frac{1}{2} (y_2 - R_2 y_1)^T \Sigma_2^{-1} (y_2 - R_2 y_1) + \text{const}
\end{align}

To maximize, differentiate with respect to $y_1$ and set the derivative to zero:

\begin{align}
    \frac{d}{dy_1} \ln P(y_1 | y_2) &= -\Sigma_1^{-1} (y_1 - R_0 y_0) + R_2^T \Sigma_2^{-1} (y_2 - R_2 y_1) = 0
\end{align}

Rewriting and solving for $y_1$:

\begin{equation}
    \left( \Sigma_1^{-1} + R_2^T \Sigma_2^{-1} R_2 \right) y_1 = \Sigma_1^{-1} R_0 y_0 + R_2^T \Sigma_2^{-1} y_2
\end{equation}

Finally, the Generalized Muon Trajectory Estimation (GMTE) solution is:

\begin{equation}\label{y_GMTE}
    y_{\text{GMTE}} = \left( \Sigma_1^{-1} + R_2^T \Sigma_2^{-1} R_2 \right)^{-1}
    \left( \Sigma_1^{-1} R_0 y_0 + R_2^T \Sigma_2^{-1} y_2 \right)
\end{equation}

Similarly, for x-z plane

\begin{equation}\label{x_GMTE}
    x_{\text{GMTE}} = \left( \Sigma_1^{-1} + R_2^T \Sigma_2^{-1} R_2 \right)^{-1}
    \left( \Sigma_1^{-1} R_0 x_0 + R_2^T \Sigma_2^{-1} x_2 \right)
\end{equation}

Within this context, the terms $\Sigma_1^{-1}$ denotes the inverse of the covariance matrix. It encompasses the variances and covariances pertaining to the lateral displacement $y_1$ and angle $\theta_1$ of the muon trajectory spanning the depths $z_0$ and $z_1$. Similarly, $\Sigma_2^{-1}$ incorporates the variances and covariances pertaining to the parameters $y_2$ and $\theta_2$ across the interval spanning from $z_1$ to $z_2$. 

The computation of the elements within the covariance matrix can be achieved by employing equations (\ref{eq:3_5}) - (\ref{eq:3_7}) as presented below. These equations incorporate the muon's depth and account for the associated energy loss.

Within equations (\ref{eq:3_5}) - (\ref{eq:3_7}), numerous terms and constants are encompassed. The term $\beta^2(z)$ denotes the squared velocity of the muon relative to the speed of light $c$, while $p^2(z)$ represents the squared momentum of the muon at the depth $z$. The empirical constants introduced by Lynch and Dahl \cite{lynch_approximations_1991} include $E_0=13.6$ MeV/c and $0.038$. Moreover, the quantity $X_0$ signifies the radiation length, which serves as a material-specific constant.

\begin{multline}
\label{eq:3_5}
\sigma_{y_1}^2 (z_0, z_1) = E_0^2 \left(1+0.038 \ln{\frac{z_1-z_0}{X_0}}\right)^2 \\ \times \int_{z_0}^{z_1} \frac{(z_1-z)^2}{\beta^2(z) p^2 (z)} \,\frac{dz}{X_0}
\end{multline}

\begin{multline}\label{eq:3_6}
\sigma_{\theta_1}^2 (z_0, z_1) = E_0^2 \left(1+0.038 \ln{\frac{z_1-z_0}{X_0}}\right)^2 \\ \times \int_{z_0}^{z_1} \frac{1}{\beta^2(z) p^2 (z)} \,\frac{dz}{X_0}
\end{multline}

\begin{multline}\label{eq:3_7}
\sigma_{y_1 \theta_1}^2 (z_0, z_1) = E_0^2 \left(1+0.038 \ln{\frac{z_1-z_0}{X_0}}\right)^2 \\ \times \int_{z_0}^{z_1} \frac{z_1-z}{\beta^2(z) p^2 (z)} \,\frac{dz}{X_0}
\end{multline}

\subsubsection*{Muon Energy Loss}

 $\frac{dz}{\beta^2(z) p^2 (z)}$ factor appears inside the scattering variance equations (Eqs. \ref{eq:3_5}-\ref{eq:3_7}), which describe how muons scatter as they travel through materials. The accuracy of this term is critical to correctly estimating the scattering covariance matrix, and hence to the trajectory reconstruction quality in $\mu$TRec. To compute these matrices accurately, one must account for the energy dependent scattering of charged particles. Due to relativistic effects, scattering displacement moments are inversely proportional to the muon momentum. The Bethe-Bloch formula describes the average energy loss per unit length as a muon travels through a material, given by:

\begin{align}
\frac{dE}{dz} &= 4\pi N_A r_e^2 m_e c^2 Z_1 \frac{Z_2}{A_2} \frac{1}{\beta^2} \nonumber \\
&\quad \times
\left[
\ln\left(
\frac{2m_e c^2 \beta^2 \gamma^2 T_{\max}}{I^2}
\right)
- \beta^2 - \frac{\delta}{2}
\right] \quad \text{(MeV/cm)},
\label{eq:bethe-bloch}
\end{align}

where $N_A$ is Avogadro’s number, $r_e$ is the classical electron radius, $m_e$ is the electron mass, $Z_1$ is the charge of the incident muon (typically 1), $Z_2$ and $A_2$ are the atomic number and mass number of the absorber medium, respectively, and $I$ is the mean excitation potential of the medium. The quantity $T_{\max}$ represents the maximum energy that can be transferred to an electron in a single collision and is given by:

\begin{equation}
T_{\max} = \frac{2m_e c^2 \beta^2 \gamma^2}{1 + 2\gamma m_e / M + (m_e / M)^2},
\label{eq:tmax}
\end{equation}

where $M$ is the muon mass and $\gamma = (1 - \beta^2)^{-1/2}$ is the Lorentz factor. However, for the energy range relevant to cosmic-ray muons, i.e. $1~\text{GeV} < E < 60~\text{GeV}$ and for most materials, radiative losses and nuclear effects are negligible. Therefore, the energy loss can be approximated as a constant:

\begin{equation}
\frac{dE}{dz} = -a,
\label{eq:constant_loss}
\end{equation}

where $a$ is a material dependent constant. For example, $a \approx 1$--$2~\text{MeV/cm}$ for most materials, ranging from $1.992~\text{MeV/cm}$ in water up to $20.5~\text{MeV/cm}$ in uranium. From special relativity:

\begin{equation}
E^2 = p^2 c^2 + m^2 c^4 \quad \Rightarrow \quad p = \frac{\sqrt{E^2 - m^2 c^4}}{c}
\end{equation}

For high-energy muons where $E \gg mc^2$, we can approximate:

\begin{equation}
p \approx \frac{E}{c} \quad \Rightarrow \quad E \approx pc
\end{equation}

In natural units where $c = 1$, this becomes $p \approx E$, allowing us to relate energy loss directly to momentum loss.

using the chain rule:

\begin{equation}
\frac{dp}{dz} = \frac{dE}{dz} \cdot \frac{dp}{dE} = -a \cdot 1 = -a
\end{equation}

Integrating both sides:

\begin{equation}
\int dp = -a \int dz \quad \Rightarrow \quad p(z) = p_0 - a z
\label{eq:momentum_loss}
\end{equation}

This expression is crucial in scattering variance calculations, where $p(z)$ replaces energy-dependent momentum in path integrals. Now, using the relativistic limit $\beta(z) \approx 1$, we obtain:

\begin{equation}
\frac{1}{\beta(z)^2 p(z)^2} \approx \frac{1}{p(z)^2} = \frac{1}{(p_0 - az)^2},
\label{eq:scattering_factor_approx}
\end{equation}

The integral used in the scattering angle variance expressions can now be evaluated:

\begin{align}\label{energy_loss}
\int \frac{dz}{\beta(z)^2 p(z)^2} &= \int \frac{dz}{(p_0 - az)^2} = \frac{z}{p_0(p_0 - az)} + C
\end{align}

where $C$ is the constant of integration. This approximation enables analytical or numerical evaluation of the scattering behavior of muons as a function of material depth and incident energy.

\section{Scattering angle mapping}

\begin{figure}[h]
    \centering
    {\includegraphics[width=245px]{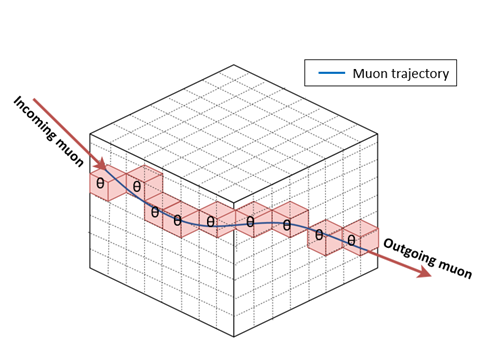}}
    \caption{Scattering angle assigned to each voxel traversed by the muon path.}
    \label{fig:density_mapping}
\end{figure}

In this analysis, each voxel traversed by a muon is assigned a specific scattering angle as shown in Figure \ref{fig:density_mapping}. The scattering angle for each muon passing through the medium is calculated using the relation:

\begin{align}\label{eq:scattering_angle}
\theta 
= \sqrt{\frac{\theta_{x}^2 + \theta_{y}^2}{2}},
\end{align} 
where the angular deviations in the $x$ and $y$ directions, $\theta_x$ and $\theta_y$, are given by:

\begin{align}\label{eq:scattering_angle_x}
\theta_{x} 
= \tan^{-1}\!\left(\frac{x_{4} - x_{3}}{z_{4} - z_{3}}\right)
- \tan^{-1}\!\left(\frac{x_{2} - x_{1}}{z_{2} - z_{1}}\right),
\end{align} 

\begin{align}\label{eq:scattering_angle_y}
\theta_{y} 
= \tan^{-1}\!\left(\frac{y_{4} - y_{3}}{z_{4} - z_{3}}\right)
- \tan^{-1}\!\left(\frac{y_{2} - y_{1}}{z_{2} - z_{1}}\right),
\end{align} 

with $(x_1, y_1, z_1)$ and $(x_2, y_2, z_2)$ representing entry coordinates, and $(x_3, y_3, z_3)$ and $(x_4, y_4, z_4)$ representing exit coordinates of the muons through the detection system.

The voxel value assigned for reconstruction purposes is computed as the average of scattering angles from all muons that intersect that particular voxel, as expressed by:

\begin{align}\label{eq:voxel_value}
\text{Voxel Value}
= \frac{1}{N} \sum_{i=1}^{N} \theta_i,
\end{align} 
where $\theta_i$ is the scattering angle of the $i^\text{th}$ muon, and $N$ is the total number of muons passing through the voxel. A pseudocode for the $\mu$TRec algorithm is shown in Fig. \ref{fig:mutrec-pseudocode}.

\begin{figure}[htbp]
\centering
\begin{minipage}{0.95\linewidth}
\begin{tcolorbox}
\begin{enumerate}
    \item Collect incident and exiting muon points: \\
    \hspace*{1em} $(p_1, p_2)$ from entry plane, \\
    \hspace*{1em} $(p_3, p_4)$ from exit plane.
    
    \item Compute scattering angles (Eqs. \ref{eq:scattering_angle_x} and \ref{eq:scattering_angle_y}): \\
    
     \item Estimate the muon trajectory using a Gaussian MCS model and Bayesian formalism:
    \begin{itemize}
        \item[(a)] Set theoretical values for radiation length, energy loss constant, and average momentum.
        \item[(b)] Apply linear energy loss model (Eq.~\ref{energy_loss}).
        \item[(c)] Compute scattering variances $\sigma^2(z)$ from MCS theory (Eqs.~\ref{eq:3_5}--\ref{eq:3_7}).
       \item[(d)] Construct covariance matrices $\Sigma_1(z), \Sigma_2(z)$ (Eq. \ref{covariance_matrices}) and transport matrices $R_1(z), R_2(z)$ (Eq. \ref{Transport_matrices}).
        \item[(e)] Use Bayesian update to compute estimated trajectory (Eqs.~\ref{y_GMTE}, \ref{x_GMTE}).
    \end{itemize}

    \item Discretize the reconstruction volume into $n \times n \times n$ voxels.
    \begin{itemize}
       \item[(a)] Assign a $\theta$ value (Eq. \ref{eq:scattering_angle}) to each voxel that muon trajectory passes through.
        \item[(b)] Repeat for all muons and average voxel statistics to form a scattering density map (Eq. \ref{eq:voxel_value}). 
    \end{itemize}
    
    \item Reconstruct image using the computed voxel statistics.
    
\end{enumerate}
\end{tcolorbox}
\end{minipage}
\caption{Pseudocode for the $\mu$TRec algorithm.}
\label{fig:mutrec-pseudocode}
\end{figure}

\section{Results}

\subsection{Mono-energetic unidirectional muons}

\begin{figure*}[htbp]
    \centering
      \label{fig:a}%
      \includegraphics[width=0.25\textwidth]{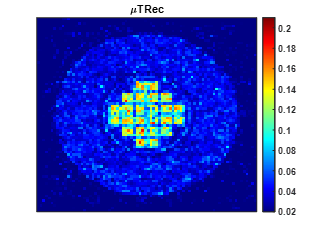}%
    \hfill
      \label{fig:b}%
      \includegraphics[width=0.25\textwidth]{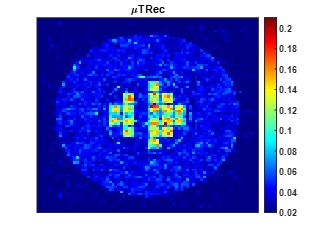}%
    \hfill
      \label{fig:c}%
      \includegraphics[width=0.25\textwidth]{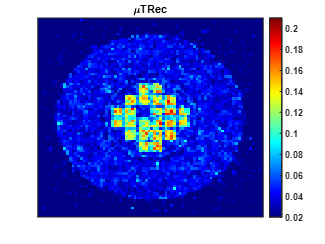}%
    \hfill
      \label{fig:d}%
      \includegraphics[width=0.25\textwidth]{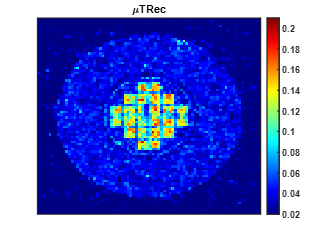}%

    
    \subfloat[]{%
      \label{fig:e}%
      \includegraphics[width=0.25\textwidth]{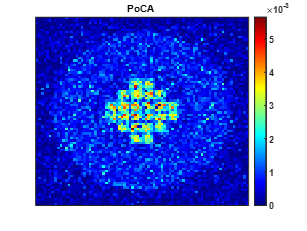}%
    }%
    \hfill
    \subfloat[]{%
      \label{fig:f}%
      \includegraphics[width=0.25\textwidth]{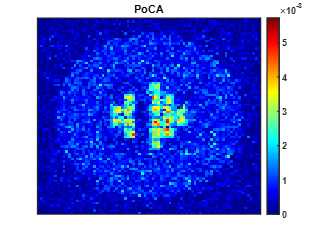}%
    }%
    \hfill
    \subfloat[]{%
      \label{fig:g}%
      \includegraphics[width=0.25\textwidth]{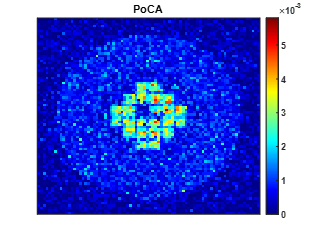}%
    }%
    \hfill
    \subfloat[]{%
      \label{fig:h}%
      \includegraphics[width=0.25\textwidth]{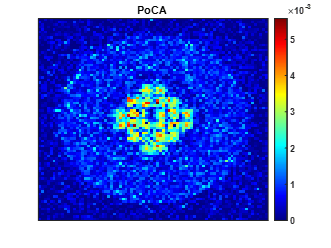}%
    }%
    
    \caption{Four configurations of the DSC reconstructed using \textbf{mono-energetic, unidirectional} $\mathbf{10^5}$ \textbf{muons} and \textbf{a voxel size of 5 cm}: (a) Fully loaded (b) One column missing (c) One fuel assembly missing and (d) Half fuel assembly missing.}
    \label{results:horizontal_100k}
\end{figure*}

\begin{figure*}[htbp]
    \centering
      \label{fig:a}%
      \includegraphics[width=0.25\textwidth]{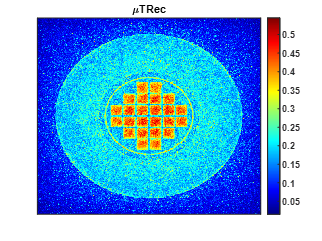}%
    \hfill
      \label{fig:b}%
      \includegraphics[width=0.25\textwidth]{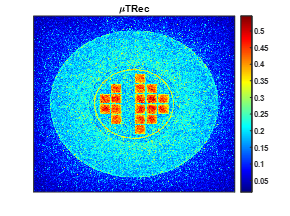}%
    \hfill
      \label{fig:c}%
      \includegraphics[width=0.25\textwidth]{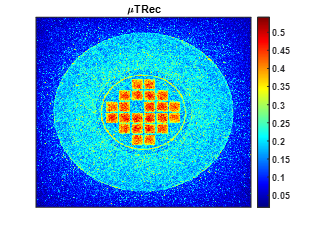}%
    \hfill
      \label{fig:d}%
      \includegraphics[width=0.25\textwidth]{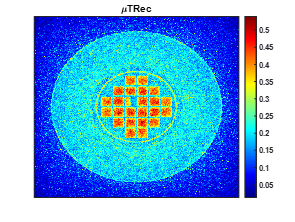}%

    
    \subfloat[]{%
      \label{fig:e}%
      \includegraphics[width=0.25\textwidth]{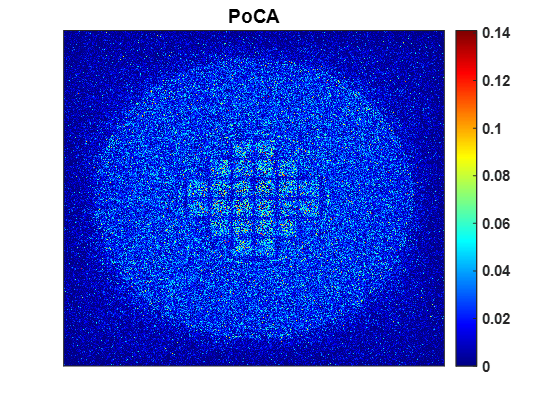}%
    }%
    \hfill
    \subfloat[]{%
      \label{fig:f}%
      \includegraphics[width=0.25\textwidth]{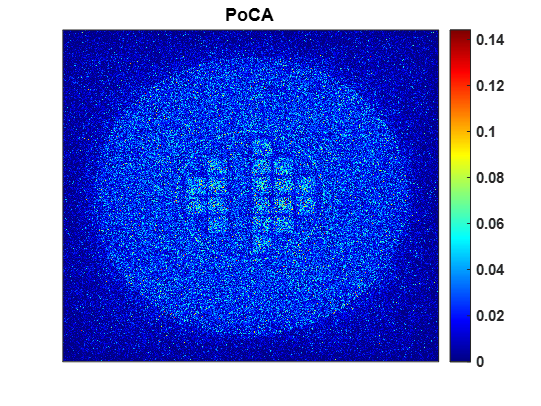}%
    }%
    \hfill
    \subfloat[]{%
      \label{fig:g}%
      \includegraphics[width=0.25\textwidth]{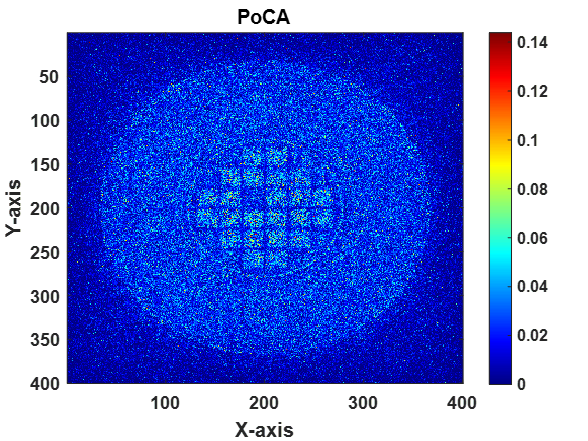}%
    }%
    \hfill
    \subfloat[]{%
      \label{fig:h}%
      \includegraphics[width=0.25\textwidth]{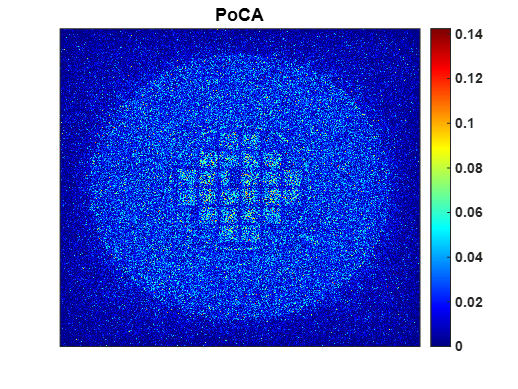}%
    }%
    
    \caption{Four configurations of the DSC reconstructed using \textbf{mono-energetic, unidirectional} $\mathbf{10^6}$ \textbf{muons} and \textbf{a voxel size of 1 cm}: (a) Fully loaded (b) One column missing (c) One fuel assembly missing and (d) Half fuel assembly missing.}
    \label{results:horizontal_1000k}
\end{figure*}

Muons with an average energy of 5~GeV, incident perpendicular to the detector planes, are simulated for the tomographic reconstruction of a horizontally oriented VSC-24 dry storage cask. A reconstruction volume of $4\times4\times4$~m$^3$ is defined, with voxel sizes of 1~cm and 5~cm corresponding to muon fluxes of $10^5$ and $10^6$, respectively. Reconstruction results are presented for four scenarios: (a) fully loaded cask, (b) one column of fuel assemblies missing, (c) one fuel assembly missing, and (d) a half fuel assembly missing, as shown in Figures~\ref{results:horizontal_100k} and~\ref{results:horizontal_1000k} for $10^5$ and $10^6$ muons, respectively. The results represent the average signal across each voxel layer along the z-axis for both PoCA and $\mu$TRec. The results demonstrate that both the image resolution and contrast improve with increasing muon flux and through the application of the $\mu$TRec algorithm. A comparative analysis of Figures~\ref{results:horizontal_100k} and~\ref{results:horizontal_1000k} indicates that the best resolution is achieved with $10^6$ muons using the $\mu$TRec approach. Notably, the $\mu$TRec algorithm is capable of identifying even subtle anomalies, such as a half-missing fuel assembly, with high spatial precision at muon fluxes as low as $10^5$.

\begin{figure}[h]
    \centering
    {\includegraphics[width=200px]{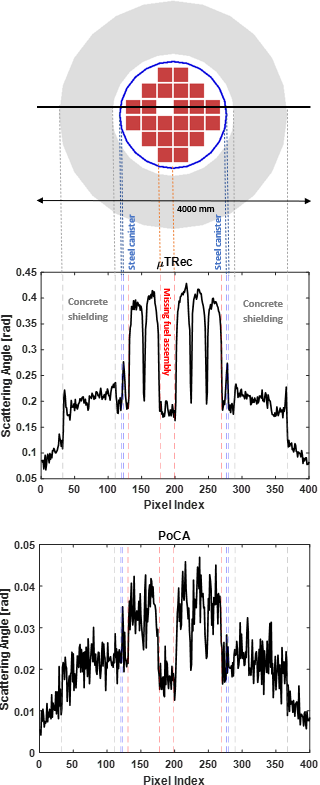}}
    \caption{Spatially averaged scattering angle signal amplitudes over a 0–210 mm y-segment (fuel assembly width) plotted along the x-direction from 0 to 4000 mm for \textbf{mono-energetic unidirectional muons with voxel size of 1 cm.}}
    \label{fig:plot_mono}
\end{figure}

Furthermore, the signal distributions for a flux of $10^6$ muons were evaluated along a horizontal line intersecting multiple material layers, including surrounding air, concrete, the air gap, and the fuel assemblies, as depicted in Figure~\ref{fig:plot_mono}. The signal amplitudes corresponding to scattering angles were spatially averaged over a 0--210~mm segment in the $y$-direction equivalent to the cross-sectional width of a single fuel assembly and plotted along the $x$-axis from 0 to 4000~mm. The resulting profiles demonstrate that the $\mu$TRec algorithm yields a more distinct material differentiation compared to the classical PoCA method. Notably, $\mu$TRec produces a smoother signal distribution and is capable of clearly resolving the absence of a single fuel assembly. In addition, the narrow gaps between individual fuel assemblies become discernible. While it is difficult for the PoCA reconstruction to clearly identify the steel canister boundaries, the $\mu$TRec algorithm successfully captures both the location and the scattering signature of the canister, further highlighting its superior spatial resolution and material sensitivity.

\subsection{Poly-energetic multidirectional muons}

\begin{figure}[h]
    \centering
    {\includegraphics[width=240px]{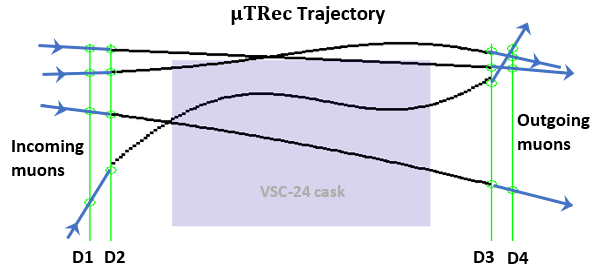}}
    \caption{Trajectories of four randomly selected muons passing through the dry storage cask, modeled using the $\mu$TRec algorithm. D1, D2, D3, and D4 denote the four scintillation detectors.}
    \label{fig:trajectory}
\end{figure}

A comparable analysis was conducted using a more realistic muon spectrum generated with the “Muon Generator,” featuring muon energies ranging from 1–60 GeV and zenith angles between 0° and 90°. The reconstruction was performed without explicit momentum information; instead, an average muon momentum of 5 GeV/c was assumed in the $\mu$TRec algorithm for calculating scattering variances. The trajectories of four randomly selected muons passing through the cask, modeled using the $\mu$TRec algorithm, are shown in Figure \ref{fig:trajectory}. It is observed that trajectories corresponding to higher scattering angles deviate slightly from the incoming and outgoing paths. This deviation may be attributed to the loss of momentum information and the assumptions made regarding the radiation length and energy loss constants.

\begin{figure*}[htbp]
    \centering
      \label{fig:a}%
      \includegraphics[width=0.25\textwidth]{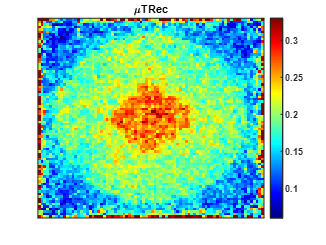}%
    \hfill
      \label{fig:b}%
      \includegraphics[width=0.25\textwidth]{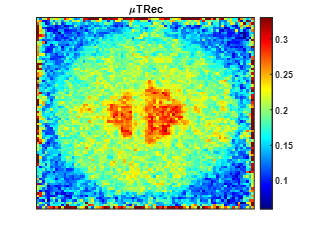}%
    \hfill
      \label{fig:c}%
      \includegraphics[width=0.25\textwidth]{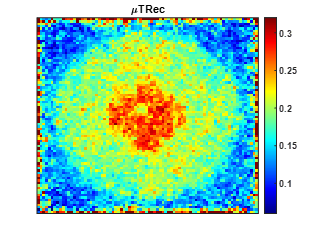}%
    \hfill
      \label{fig:d}%
      \includegraphics[width=0.25\textwidth]{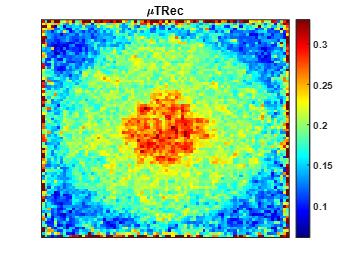}%

    
    \subfloat[]{%
      \label{fig:e}%
      \includegraphics[width=0.25\textwidth]{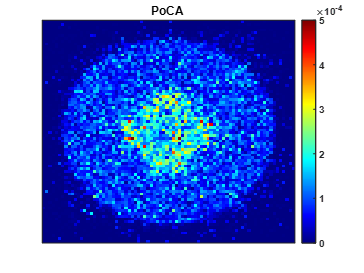}%
    }%
    \hfill
    \subfloat[]{%
      \label{fig:f}%
      \includegraphics[width=0.25\textwidth]{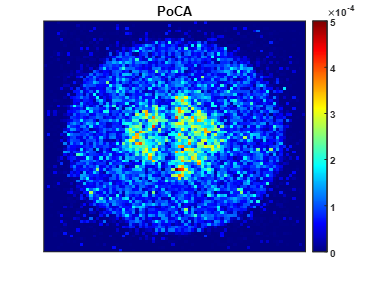}%
    }%
    \hfill
    \subfloat[]{%
      \label{fig:g}%
      \includegraphics[width=0.25\textwidth]{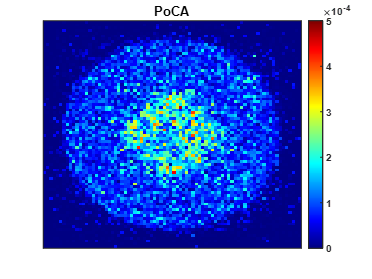}%
    }%
    \hfill
    \subfloat[]{%
      \label{fig:h}%
      \includegraphics[width=0.25\textwidth]{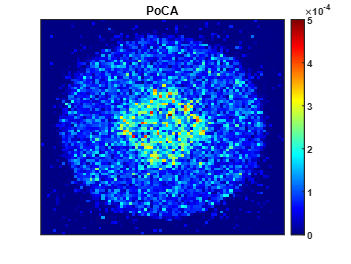}%
    }%
    
    \caption{Four configurations of the DSC reconstructed using \textbf{realistic muon spectrum} with $\mathbf{10^5}$ \textbf{muons} and \textbf{a voxel size of 1 cm}: (a) Fully loaded (b) One column missing (c) One fuel assembly missing and (d) Half fuel assembly missing.}
    \label{results:poly_100k}
\end{figure*}

\begin{figure*}[htbp]
    \centering
      \label{fig:a}%
      \includegraphics[width=0.25\textwidth]{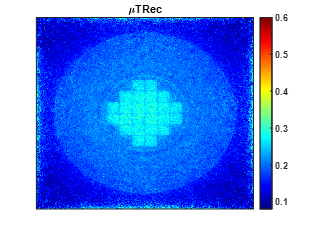}%
    \hfill
      \label{fig:b}%
      \includegraphics[width=0.25\textwidth]{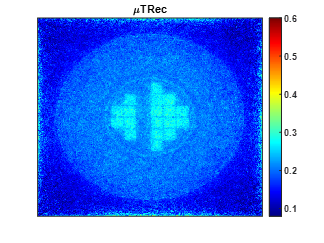}%
    \hfill
      \label{fig:c}%
      \includegraphics[width=0.25\textwidth]{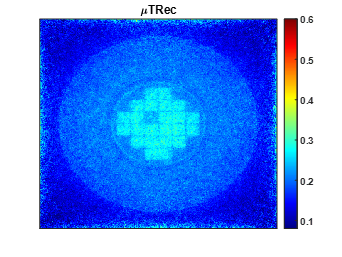}%
    \hfill
      \label{fig:d}%
      \includegraphics[width=0.25\textwidth]{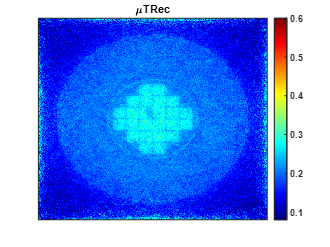}%

    
    \subfloat[]{%
      \label{fig:e}%
      \includegraphics[width=0.25\textwidth]{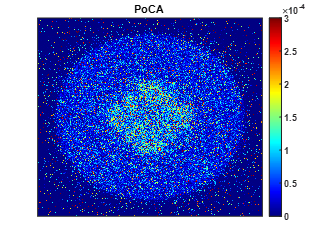}%
    }%
    \hfill
    \subfloat[]{%
      \label{fig:f}%
      \includegraphics[width=0.25\textwidth]{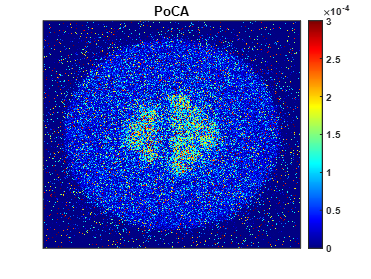}%
    }%
    \hfill
    \subfloat[]{%
      \label{fig:g}%
      \includegraphics[width=0.25\textwidth]{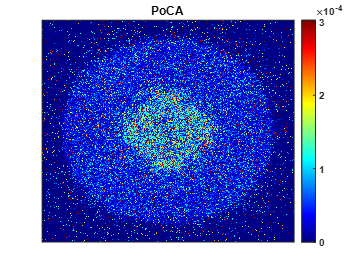}%
    }%
    \hfill
    \subfloat[]{%
      \label{fig:h}%
      \includegraphics[width=0.25\textwidth]{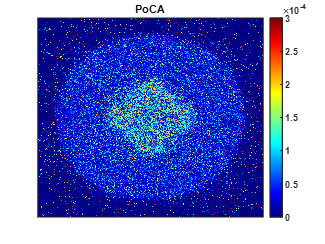}%
    }%
    
    \caption{Four configurations of the DSC reconstructed using \textbf{realistic muon spectrum} with $\mathbf{10^6}$ \textbf{muons} and \textbf{a voxel size of 1 cm}: (a) Fully loaded (b) One column missing (c) One fuel assembly missing and (d) Half fuel assembly missing.}
    \label{results:poly_1000k}
\end{figure*}

The resulting reconstructions for muon fluxes of $10^5$ and $10^6$ are presented in Figures~\ref{results:poly_100k} and \ref{results:poly_1000k}, respectively. For $10^5$ muons, the PoCA algorithm is capable of identifying a missing column of fuel assemblies but fails to detect cases where a single fuel assembly or half of one is absent. Moreover, the reconstructed images of a fully loaded DSC appear visually indistinguishable from half and single missing fuel assembly. In contrast, the $\mu$TRec algorithm can successfully differentiate between a fully loaded cask, a single column missing, and a single assembly missing. However, even $\mu$TRec is unable to clearly resolve the case of a half fuel assembly missing at this flux level. At higher flux, using $10^6$ muons, the $\mu$TRec algorithm significantly improves the image quality and resolution. All four scenarios with fully loaded, one column missing, one assembly missing, and half assembly missing are successfully identified. Notably, the $\mu$TRec not only detects the presence and location of the half-missing assembly but also reveals which specific half is absent. Additionally, the outline of the steel canister becomes discernible. In contrast, PoCA continues to suffer from insufficient muon statistics, resulting in inadequate signal strength for confident identification. While attempts were made to enhance PoCA performance by filtering high scattering angle events, only the case of a missing column was marginally distinguishable.

\begin{figure*}[htbp]
    \centering
      \label{fig:a}%
      \includegraphics[width=0.25\textwidth]{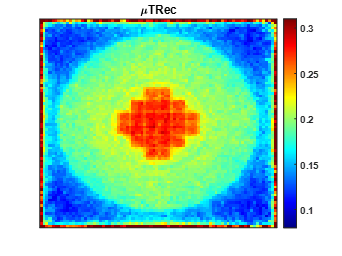}%
    \hfill
      \label{fig:b}%
      \includegraphics[width=0.25\textwidth]{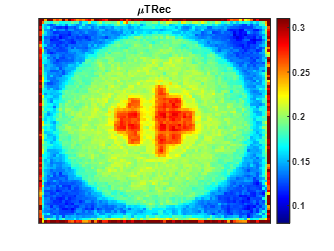}%
    \hfill
      \label{fig:c}%
      \includegraphics[width=0.25\textwidth]{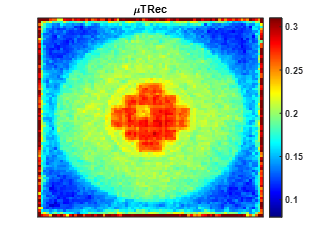}%
    \hfill
      \label{fig:d}%
      \includegraphics[width=0.25\textwidth]{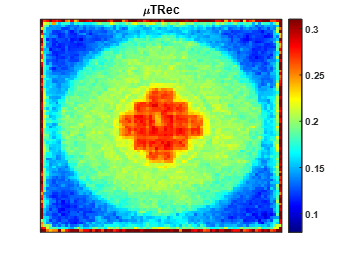}%

    
    \subfloat[]{%
      \label{fig:e}%
      \includegraphics[width=0.25\textwidth]{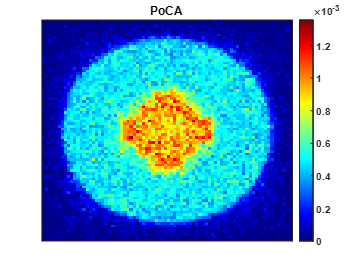}%
    }%
    \hfill
    \subfloat[]{%
      \label{fig:f}%
      \includegraphics[width=0.25\textwidth]{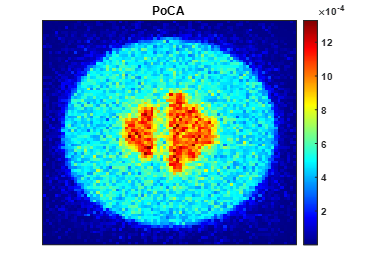}%
    }%
    \hfill
    \subfloat[]{%
      \label{fig:g}%
      \includegraphics[width=0.25\textwidth]{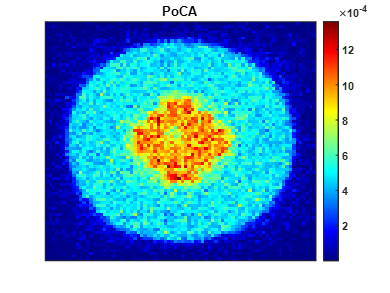}%
    }%
    \hfill
    \subfloat[]{%
      \label{fig:h}%
      \includegraphics[width=0.25\textwidth]{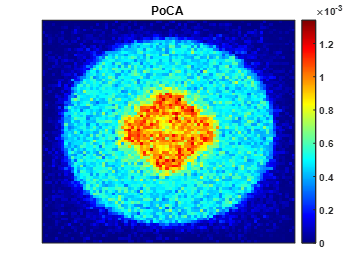}%
    }%
    
    \caption{Four configurations of the DSC reconstructed using \textbf{realistic muon spectrum} with $\mathbf{10^6}$ \textbf{muons} and \textbf{a voxel size of 5 cm}: (a) Fully loaded (b) One column missing (c) One fuel assembly missing and (d) Half fuel assembly missing.}
    \label{results:poly_1000k_5cm}
\end{figure*}

To address the limitations imposed by low signal strength at smaller voxel sizes, the voxel size was increased to 5 cm for the simulation involving $10^6$ muons. At this spatial resolution, the number of muons become sufficient to generate meaningful signals for the PoCA algorithm when higher scattering angle events are excluded. As shown in Figure~\ref{results:poly_1000k_5cm}, the signal amplitude improves for both PoCA and $\mu$TRec. For the PoCA-based reconstruction, the increased voxel size enables successful identification of the scenarios with one missing column and one missing fuel assembly. However, the algorithm continues to struggle in localizing the half-missing fuel assembly, and it remains inconclusive in confirming that the fully loaded configuration contains no missing fuel assemblies. In contrast, the $\mu$TRec algorithm maintains high fidelity even at the coarser resolution. Although the outline of the steel canister becomes less distinct due to its thickness (2.5 cm) being smaller than the voxel size, the overall signal strength improves. $\mu$TRec clearly distinguishes all four fuel-loading configurations, including the half fuel assembly missing scenario.

\begin{figure}[h]
    \centering
    {\includegraphics[width=200px]{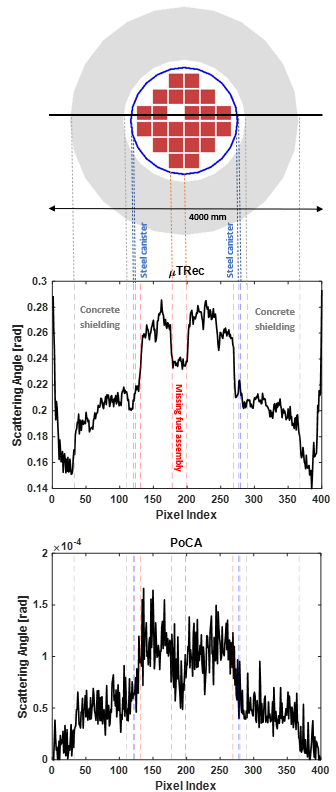}}
    \caption{Spatially averaged scattering angle signal amplitudes over a 0–210 mm y-segment (fuel assembly width) plotted along the x-direction from 0 to 4000 mm for \textbf{realistic muon energy spectrum with voxel size of 1 cm.}}
    \label{fig:plot_poly}
\end{figure}

\begin{figure}[h]
    \centering
    {\includegraphics[width=200px]{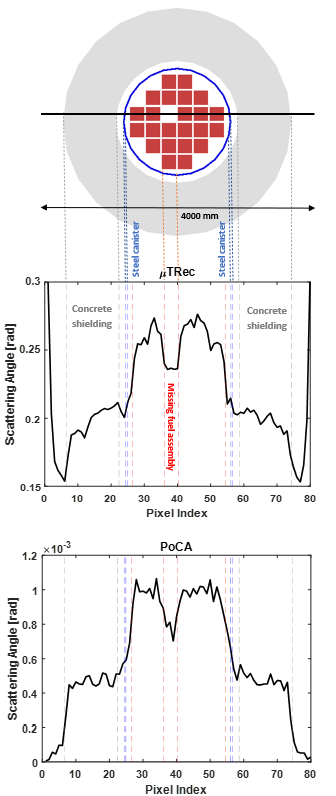}}
    \caption{Spatially averaged scattering angle signal amplitudes over a 0–210 mm y-segment (fuel assembly width) plotted along the x-direction from 0 to 4000 mm for \textbf{realistic muon energy spectrum with voxel size of 5 cm.}}
    \label{fig:plot_poly_5cm}
\end{figure}

The evaluation of signal distributions for a flux of \(10^6\) muons generated using a realistic cosmic ray spectrum is presented in Figure~\ref{fig:plot_poly}, where a horizontal profile intersecting multiple material layers is analyzed with a voxel size of 1~cm. The results clearly demonstrate the superior material differentiation capability of the $\mu$TRec algorithm compared to the conventional PoCA method. $\mu$TRec provides a smoother and more continuous signal distribution, enabling clear identification of the absence of a single fuel assembly. In addition, fine structural features such as inter-assembly gaps and the boundaries of the steel canister are distinctly resolved. In contrast, the PoCA algorithm exhibits a significantly weaker signal, making it difficult to even detect the missing fuel assembly. Figure~\ref{fig:plot_poly_5cm} shows the same analysis with an increased voxel size of 5~cm. While the larger voxel improves the signal strength for PoCA, enabling the detection of a missing fuel assembly, the method still underperforms relative to $\mu$TRec in terms of spatial resolution and accuracy. Tables~\ref{table:10^5} and \ref{table:10^6} provide a quantitative comparison between PoCA and $\mu$TRec in terms of signal-to-noise ratio (SNR), contrast-to-noise ratio (CNR), and detection power (DP) for the case of a single missing fuel assembly under \(10^5\) and \(10^6\) muon fluxes, respectively \cite{liu_muon_2018}. These metrics are calculated as follows:

\begin{align}
\mathrm{SNR} &= \frac{\mathrm{mean}(\text{8 assemblies surrounding missing one})}{\mathrm{std}(\text{8 assemblies surrounding missing one})} \\
\mathrm{CNR} &= \frac{\mathrm{mean}(\text{8 assemblies}) - \mathrm{mean}(\text{missing one})}{\max(\mathrm{std}(\text{8 assemblies}), \mathrm{std}(\text{missing one}))} \\
\mathrm{DP}  &= \mathrm{SNR} \times \mathrm{CNR}
\end{align}

Here, \textit{std} denotes the standard deviation. Two regions of interest are considered: (1) the missing fuel assembly and (2) the eight surrounding fuel assemblies. The SNR quantifies the strength and uniformity of the signal within a given region. The CNR indicates the distinguishability between the two regions with higher values reflecting better material differentiation. The product of SNR and CNR is defined as the DP which serves as an overall metric for detection performance. For a muon flux of \(10^5\), $\mu$TRec outperforms PoCA significantly, with SNR, CNR, and DP improving by 357\%, 488\%, and 2593\%, respectively. These large improvements are attributed to the insufficient muon statistics in PoCA which prevent reliable detection, while $\mu$TRec remains effective even at lower fluxes. At a higher flux of \(10^6\) muons and a voxel size of 1~cm, $\mu$TRec achieves even greater enhancements: SNR, CNR, and DP increase by 1147\%, 603\%, and 8546\%, respectively, compared to PoCA. When the voxel size is increased to 5~cm, PoCA performance improves due to increased signal strength making a more direct comparison with $\mu$TRec possible albeit at the cost of reduced resolution in $\mu$TRec. In this configuration, the steel canister becomes less discernible. Nevertheless, $\mu$TRec continues to outperform PoCA, with respective improvements of 122\% in SNR, 35\% in CNR, and 201\% in DP.

\begin{table}[h]
\centering
\caption{Reconstructed image characteristics using a realistic energy spectrum of $10^5$ muons for one missing fuel assembly.}
\begin{tabular}{|p{2cm}|p{2cm}|p{2cm}|p{2cm}|}
\hline
\textbf{} & \textbf{SNR} & \textbf{CNR} & \textbf{DP} \\
\hline
\textbf{$\mu$TRec} & 12.40 & 1.55 & 19.23 \\
\textbf{PoCA} & 2.49 & 0.17 & 0.42 \\
\hline
\end{tabular}
\label{table:10^5}
\end{table}

\begin{table}[htbp]
\centering
\caption{Reconstructed image characteristics using a realistic energy spectrum of $10^6$ muons for one missing fuel assembly.}
\begin{tabular}{|p{1.8cm}|p{1.5cm}|p{1.5cm}|p{1.5cm}|p{1.5cm}|}
\hline
\textbf{Voxel size} & \textbf{Algorithm} & \textbf{SNR} & \textbf{CNR} & \textbf{DP} \\
\hline
\textbf{1 cm} & \textbf{$\mu$TRec} & 16.84 & 2.11 & 35.45 \\
 & \textbf{PoCA} & 1.35 & 0.30 & 0.41 \\
\hline
\textbf{5 cm} & \textbf{$\mu$TRec} & 19.30 & 1.85 & 35.77 \\
 & \textbf{PoCA} & 8.69 & 1.37 & 11.90 \\
\hline
\end{tabular}
\label{table:10^6}
\end{table}

\section{Conclusions}

For a flux of 10$^5$ mono-energetic, unidirectional muons, both PoCA and $\mu$TRec were able to reconstruct the general structure of the DSC and detect larger anomalies such as one column or one full fuel assembly missing. However, PoCA struggled to reliably identify smaller defects, such as a half-missing fuel assembly. In contrast, $\mu$TRec produced a significantly stronger and smoother signal, improving image contrast and resolution. Notably, $\mu$TRec was able to localize the half-missing fuel assembly with higher confidence, and clearly delineate individual fuel assemblies and the narrow gaps between them, demonstrating its advantage even at low muon statistics. At a higher flux of 10$^6$ muons, image quality improved for both PoCA and $\mu$TRec, but the performance gap between the two methods widened. $\mu$TRec delivered sharper reconstructions, more accurate material boundary definitions, and higher contrast in scattering signatures. It successfully detected all four configurations with high spatial resolution, including subtle features like the narrow spacing between fuel assemblies and the presence of the steel canister. PoCA demonstrated a slight improvement in spatial resolution and sensitivity to the steel canister boundaries; however, the number of muon events were insufficient to produce signal strength as sharp as that achieved with the $\mu$TRec algorithm.

For the case of a realistic muon energy spectrum, at a flux level of 10$^5$ muons, the $\mu$TRec algorithm can successfully detect all scenarios except the case of a half-missing fuel assembly. In contrast, the PoCA method is only able to identify the absence of one column of fuel assemblies and fails to discern more subtle anomalies due to insufficient signal strength. At a higher flux level of 10$^6$ muons and a voxel size of 1~cm, $\mu$TRec demonstrates high performance by resolving all four configurations including the detection of a half-missing fuel assembly and the clear outline of the steel canister. Conversely, PoCA struggles due to an inadequate number of muons for this spatial resolution. The case of one missing column becomes detectable only after filtering out events with large scattering angles. To enhance PoCA's effectiveness, the voxel size is increased from 1~cm to 5~cm for 10$^6$ muons. At this coarser resolution, the muon count becomes sufficient to generate a meaningful signal for PoCA enabling the detection of one missing column and one missing fuel assembly. However, distinguishing between a fully loaded cask and a half-missing fuel assembly remains challenging. $\mu$TRec on the other hand maintains high reconstruction fidelity even with the increased voxel size. While the steel canister’s outline becomes less pronounced owing to its 2.5~cm thickness being smaller than the voxel resolution, the overall signal strength improves preserving $\mu$TRec's superior detection capability.

 For muon fluxes of 10$^5$ and 10$^6$, the $\mu$TRec algorithm significantly outperforms the classical PoCA method in terms of SNR, CNR and DP. The substantial improvements observed with $\mu$TRec are primarily attributed to its robustness in low-flux scenarios where PoCA suffers from inadequate muon statistics that hinder reliable detection especially when employing smaller voxel sizes. At a flux of 10$^6$ muons increasing the voxel size to 5~cm enhances the signal strength available to PoCA thereby enabling a more meaningful comparison with $\mu$TRec. However, this comes at the expense of spatial resolution particularly for $\mu$TRec as the larger voxel size reduces its ability to delineate fine structural details such as the steel canister which has a thickness of only 2.5~cm. Despite this limitation $\mu$TRec continues to demonstrate superior performance achieving respective improvements of \textbf{122\%} in SNR, \textbf{35\%} in CNR, and \textbf{201\%} in DP compared to PoCA under these conditions.

\section{Future work}

The current analysis is performed without incorporating individual muon momentum information. A logical progression of this work involves integrating muon momentum into the $\mu$TRec framework to improve the accuracy of muon trajectory. Additionally, the use of the \emph{M-value}, a parameter derived from the scattering angle and muon momentum, will be explored for material density mapping. This approach has recently demonstrated enhanced performance when applied to the PoCA algorithm~\cite{bae_fieldable_2022, bae_momentum_2024, bae_momentum-dependent_2022, bae_novel_2022, bae_image_2024}. Extending this methodology to $\mu$TRec is expected to yield further improvements in spatial resolution and material differentiation.

\section*{Supplementary material}

MATLAB script for $\mu$TRec and PoCA algorithms is publicly available at \url{https://github.com/rughade/muTRec-algorithm}.

\begin{acknowledgments}

The present research is being conducted with the support of funding provided by the Purdue University School of Nuclear Engineering and the Purdue Research Foundation. 
\end{acknowledgments}

\bibliographystyle{IEEEtran}
\bibliography{aipsamp.bbl}

\end{document}